\documentclass[%
 reprint,
superscriptaddress,
 amsmath,amssymb,
 aps,
pra,
]{revtex4-2}
\usepackage[dvipsnames]{xcolor}
\usepackage{graphicx}
\usepackage{dcolumn}
\usepackage{bm}%
\usepackage{layouts}
\usepackage{mathtools}
\usepackage{braket}
\usepackage{color}
\usepackage{xcolor}
\usepackage{graphicx}
\usepackage{dcolumn}
\usepackage{bm}
\usepackage{hyperref}

\newcommand{\berkeleyphy}{Department of Physics, University of California, Berkeley, California 94720}
\newcommand{\CIQC}{Challenge Institute for Quantum Computation, University of California, Berkeley, California 94720}
\newcommand{\LBL}{Materials Sciences Division, Lawrence Berkeley National Laboratory, Berkeley, California 94720}

\begin{document}

\preprint{APS/123-QED}

\title{Autler-Townes Spectroscopy of a Rydberg Ladder\\}

\author{Tai Xiang}
\affiliation{\berkeleyphy}
\affiliation{\CIQC}

\author{Yue-Hui Lu}
\affiliation{\berkeleyphy}
\affiliation{\CIQC}

\author{Jacquelyn Ho}
\affiliation{\berkeleyphy}
\affiliation{\CIQC}

\author{Tsai-Chen Lee}
\affiliation{\berkeleyphy}
\affiliation{\CIQC}

\author{Zhenjie Yan}
\affiliation{\berkeleyphy}
\affiliation{\CIQC}

\author{Dan M. Stamper-Kurn}
\email[]{dmsk@berkeley.edu}
\affiliation{\berkeleyphy}
\affiliation{\CIQC}
\affiliation{\LBL}

\date{\today}

\begin{abstract}
Ladder-type two-photon excitation of an atom from a ground state $\ket{g}$, to an intermediate excited state $\ket{e}$, and, finally, to a Rydberg state $\ket{r}$, has a variety of uses from quantum information to sensing.  A common scheme for detecting this transition optically is through electromagnetically induced transparency (EIT).  However, in inverted wavelength schemes, where the ground-to-excited transition wavelength is shorter than the excited-to-Rydberg transition wavelength, the strength of the EIT feature on the lower-leg beam is strongly reduced in a Doppler-broadened medium.  Here, we report on an alternative two-photon spectroscopic feature, which we term the two-photon Autler-Townes resonance, observed on the upper-leg beam.  Compared to the EIT signal, this feature's superior signal-to-noise ratio allows one to resolve Rydberg resonances with principal quantum number as high as $n=80$.  We also show that such a feature can be utilized to generate an error signal for stabilizing the frequency of the upper-leg beam.
\end{abstract}

\maketitle
\section{Introduction}

Rydberg atoms, in which an electron is excited to a state with high principal quantum number, have become an important tool for quantum sensing \cite{sensingrev1,sensingrev2,Wu_2021}, quantum simulation \cite{quantumsim,quantumsim2,Browaeys2020}, and quantum computing \cite{Wintersperger2023,Saffman_2016,pritchard_2013, saffmanrev}. Oftentimes, excitations to Rydberg states are realized via a two-photon transition, in which a first photon transfers a valence electron to an intermediate state on a lower-leg transition, and a second photon drives excitations to the Rydberg state on an upper-leg transition. In such two-step excitations, ladder-type electromagnetically induced transparency (EIT) can be observed via probing the lower-leg light \cite{Finkelstein_2023,harris_1997,pritchard_2013}.  This EIT signal can be used for locating and stabilizing two-photon resonance frequencies \cite{ferlaino_2021,xu_2025, Cheng2017,bog_2004,Ying_2014}, sensing electric fields \cite{simons2021,Feng2025}, studying many-body quantum systems \cite{Kim2021,Zhang2025}, and quantum networking \cite{Srakaew2023}.

In a Doppler-broadened medium, such as a thermal vapor, atoms from a broad range of velocity classes can be Doppler-shifted to satisfy the two-photon resonance condition.  The overall optical response of the medium is obtained as the summed response over the atomic velocity distribution.  When the wavelength of the lower-leg photon is shorter than that of the upper-leg photon (the so-called inverted wavelength scheme), the EIT feature is strongly suppressed due to single-photon absorption features from difference velocity groups smearing out the coherent EIT transmission \cite{shepherd_1996, boon_1999, Zhang2024}. 

In this paper, we identify an alternative spectroscopic feature that can be observed via probing the upper-leg (longer wavelength, in the inverted scheme) light, which we term the two-photon Autler-Townes resonance (TPAT).  We find that while Doppler broadening, atom number fluctuations and a single-photon absorption background limit the signal-to-noise ratio of the EIT feature, the TPAT resonance does not suffer from these drawbacks. We benchmark the TPAT feature for various Rydberg states and compare it against the EIT feature. Finally, we demonstrate that the TPAT feature can be applied to generate an error signal that can be used to lock the laser driving the upper-leg transition.

\section{Probing the lower-leg light: EIT}
\begin{figure*}
\includegraphics{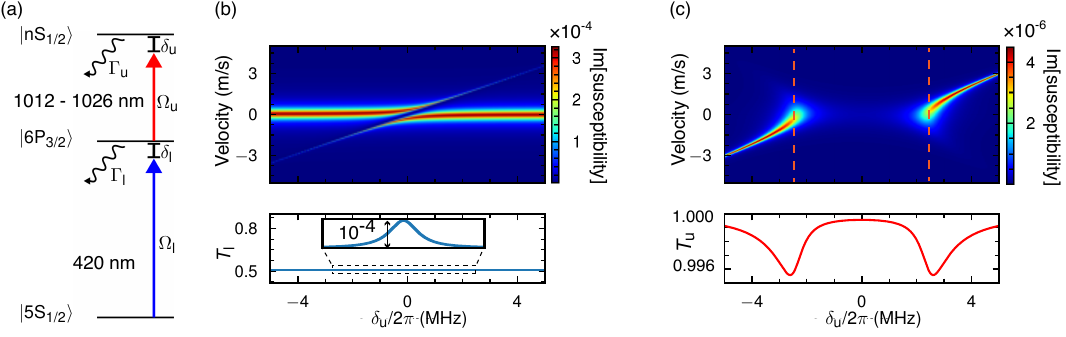}
\caption{(a) Effective three level model. 420~nm light with Rabi frequency $\Omega_{\mathrm{l}}$ couples the ground and intermediate state. 1012--1026~nm light couple the intermediate state to a range of Rydberg states with Rabi frequency $\Omega_{\mathrm{u}}$. The intermediate state decays with rate $\Gamma_{\mathrm{l}}$, and the Rydberg state decays with rate $\Gamma_{\mathrm{u}}$. (b) Simulated absorption of weak lower-leg light when the detuning of the strong upper-leg driving light is scanned (top) and resultant transmission $T_\mathrm{l}$ (bottom) obtained via integrating over all velocity classes. (c) Simulated absorption of the weak upper-leg light with strong lower-leg driving light (top) and resultant transmission $T_\mathrm{u}$ (bottom) obtained via integrating over all velocity classes and scanning $\delta_u$. The dashed lines in the top plot denote tangential cuts that can be drawn at the turning point of the signal. Parameters used: $n=30$, $\Gamma_{\mathrm{l}} = 2\pi \times 1.4$~MHz, $\Gamma_{\mathrm{u}} = 2\pi\times11$~kHz,  $T=89^\circ$C; $\Omega_{\mathrm{l}} = 2\pi \times 40$~kHz and $\Omega_{\mathrm{u}} = 2\pi \times 1.2$~MHz in (b); $\Omega_{\mathrm{l}} = 2\pi \times 4.8$~MHz and  $\Omega_{\mathrm{u}} = 2\pi \times36$~kHz in (c). Atomic values obtained from the Alkaline Rydberg Calculator \cite{arc}. }\label{fig:sys_theory}
\end{figure*}

Let us examine the effective three-level system formed by the ground state $\ket{g}$, the intermediate state $\ket{e}$, and the Rydberg state $\ket{r}$. In later simulation and its comparison to experiment, we will specifically consider $^{87}$Rb atoms and map the states as $\ket{g} = \ket{5S_{1/2},\text{ }F=2}$, $\ket{e} = \ket{6P_{3/2},\text{ }F'=3}$, and $\ket{r} = \ket{30S_{1/2}}$, see Fig. \ref{fig:sys_theory}(a).  
In such a system, one can probe the two-photon resonance by coupling $\ket{g}$ to $\ket{e}$ with weak lower-leg (probe) light, and $\ket{e}$ to $\ket{r}$ with strong upper-leg (dressing) light, and then measuring the transmission properties of the lower-leg light.  The strong upper-leg light dresses the intermediate state, and the weak lower-leg light couples to both of the dressed intermediate states.  This coupling results in three prominent features as a function of the lower-leg light frequency: resonant absorption away from the two-photon resonance on each of the dressed-state resonances (i.e.\ Autler-Townes absorption features), and EIT right at the two-photon resonance and between the Autler-Townes absorption resonances.  This EIT can be interpreted as coming about from destructive interference of transition amplitudes from the ground state on each of the dressed states.  Equivalently, EIT can be understood as the signature of a stationary dark state in the three-level dressed atomic system, comprised of coherent population in the ground and Rydberg state, for which transitions to the intermediate state from both the ground state (driven by lower-leg light) and the Rydberg state (driven by upper-leg light) interfere destructively.  This destructive interference reduces the absorption of the lower-leg light \cite{harris_1997}.

Importantly, the two-photon conditions that lead to lower-leg absorption (at the Autler-Townes resonances) and transparency (EIT, precisely at the two-photon resonance) are velocity dependent. For counter-propagating lower- and upper-leg light beams, the two-photon resonance condition is satified for atoms at velocity $\mathbf{v}$ whenever $(\delta_\mathrm{l} - \mathbf{k}_\mathrm{l} \cdot \mathbf{v}) - (\delta_\mathrm{u} + \mathbf{k}_\mathrm{u} \cdot \mathbf{v})  = 0$, where $\delta_{\mathrm{l},\mathrm{u}}$ are the bare lower- and upper-leg detunings, and $\mathbf{k}_{\mathrm{l},\mathrm{u}}$ the lower- and upper-leg wavevectors.

In a hot medium, such as a room-temperature thermal atomic vapor, the Doppler shifts $\mathbf{k}_\mathrm{l} \cdot \mathbf{v}$ and $\mathbf{k}_\mathrm{u} \cdot \mathbf{v}$ can be far larger than the widths of the absorption and EIT features.  In this case, and for the inverted-ladder scheme, at near-resonant settings of the driving light beams (i.e.\ with respect to the resonances for a stationary atom), all three spectroscopic features are overlaid: for atoms at three different velocity classes, one achieves the resonance conditions on each of the two Autler-Townes and also for the EIT condition. We simulate this phenomenon using the three level ladder model shown in Fig. \ref{fig:sys_theory}(a), incorporating spontaneous emission from the intermediate and Rydberg state via the Lindblad master equation. Doing so yields the overlay shown in Fig.\ \ref{fig:sys_theory}(b). The simulated absorption profiles are obtained from the imaginary part of the optical susceptibility, which represents the dissipative response of the atomic medium and is directly proportional to the lower-leg beam absorption. 

For example, focusing on the condition where both the lower-leg and upper-leg light are exactly resonant with the transition frequencies for a stationary atom ($\delta_\mathrm{l}$ = $\delta_\mathrm{u}$ = 0), the lower-leg light is simultaneously affected by Autler-Townes absorption for atoms with positive and negative velocities along the $\mathbf{k}_\mathrm{l}$ direction, and also by full EIT for the atoms at zero velocity.  The net effect of absorption by all velocity classes largely covers up the EIT signal that would otherwise be observed clearly in a cold medium with negligible Doppler shifts.  The fact that the Autler-Townes absorption for non-zero-velocity atoms tends to cover up the EIT that affects zero-velocity atoms at this setting is a consequence of the wavelength relation for the lower- and upper-leg transitions in the inverted scheme, as has been explained previously \cite{shepherd_1996, boon_1999, Zhang2024} and observed in several experiments \cite{xu_2025, Chen_2020, Urvoy_2013,Xu2016}. In contrast, in the non-inverted scheme in which the lower-leg transition has a longer wavelength, the Doppler shifts on the lower- and upper-leg transitions lead the Autler-Townes absorption features to veer away from one another. As a result, they do not obscure the zero-velocity EIT signature, leading to robust EIT spectroscopic features that persist in Doppler-broadened media \cite{Zhao:09,Simons_2016}.

While the visibility of the EIT feature in a Doppler-broadened medium is reduced in the inverted excitation scheme, it is not suppressed altogether.  The residual fractional variation in the optical density of the atomic medium depends on several factors, including the finite Doppler width of the medium, the linewidths of the two excited states ($|e\rangle$ and $|r\rangle$) being probed, the presence of dark states to which the optically driven atom can decay, and saturation effects as the intensity of the lower- and upper-leg light beams are varied. Such effects have been previously examined in Refs.\ \cite{Wu2017,Chen_2020,Moon2008,Su2022}, and we observe similar effects in our system that enhance the EIT magnitude beyond what is shown in Fig. \ref{fig:sys_theory}(b). Additionally, under certain conditions, the zero-velocity two-photon resonance can be seen not as an EIT-generated decrease in absorption, but rather as a slight \emph{increase} in optical density \cite{Urvoy_2013}.

\section{Probing the upper-leg light: TPAT}

An alternative for optically detecting two-photon resonances is to probe the upper-leg transition. We consider a strong lower-leg light beam coupling $\ket{g}$ and $\ket{e}$ and a weak upper-leg light beam coupling $\ket{e}$ and $\ket{r}$. The strong lower-leg light splits the intermediate state into two dressed states separated by the lower-leg Rabi frequency. 
Two-photon resonance is achieved when the upper-leg light is detuned sufficiently both above and below the bare atom upper-leg resonance frequency. At these light-shifted resonances, the upper-leg light is absorbed, appearing as an Autler–Townes doublet when measuring its transmission. We term this feature to be the two-photon Autler-Townes (TPAT) signal. Similar Autler–Townes–type spectra, including the effects of Doppler broadening and velocity averaging in thermal atomic vapors, have been investigated previously in diamond-type four-wave mixing configurations \cite{Whiting2017_FWM_HPB, Carr2012_RydbergFWM, Parniak2015_DiamondFWM}.

As before, we now consider the absorption signal generated by all velocity classes in the Doppler-broadened atomic ensemble being probed; see Fig.\  \ref{fig:sys_theory}(c). Here, the two bright branches on our plot correspond to the velocity classes that contribute to two-photon absorption. 
In the inverted wavelength case, the combination of single-photon and two-photon detunings yield two-photon-resonant velocity class branches that curve away from each other, in contrast to the lower-leg EIT case, where the branches overlap.
As a result, an avoided crossing is strongly resolved and a tangential cut can be drawn at the turning point of the curvature, as shown in Fig. \ref{fig:sys_theory}(c), meaning a larger number of velocity classes can contribute to the transition and enhance the absorption of the upper-leg light. 

\section{Experimental Setup}\label{sec:setup}
\begin{figure}[!htbp]
\includegraphics[width=\linewidth]{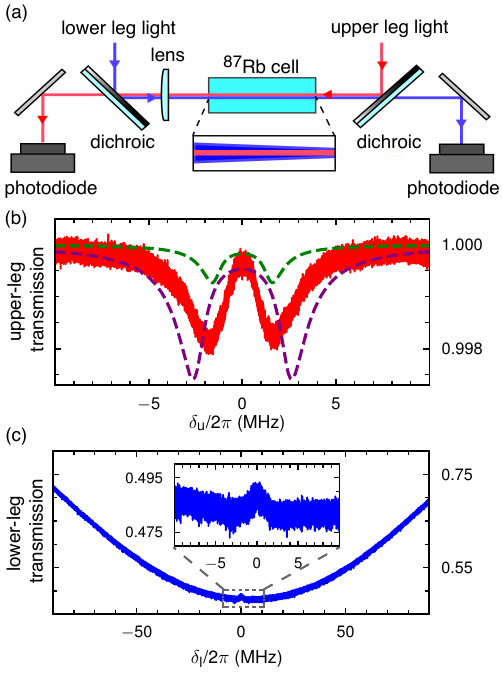}
\caption{(a) Overview of the experimental apparatus. Inset denotes the profile of the lower-leg beam (blue)
and the upper-leg beam (red). (b) Measured upper-leg transmission $T_\mathrm{u}$ with TPAT signal. The lower-leg beam is resonant with the $^{87}$Rb $\ket{5S_{1/2},\text{ }F=2} \rightarrow \ket{6P_{3/2}, \text{ }F'=3}$ hyperfine transition at zero velocity. The upper-leg beam is scanned across the two-photon resonance of the Rydberg state. The dashed purple (green) line is the simulated transmission using the experimental parameters when the ground state atoms are spin-polarized in the $^{87}$Rb $\ket{5S_{1/2},\text{ }F=2, \text{ } m_F = 2}$ sub-level (unpolarized), the lower-leg light is $\sigma^{+}$ polarized, and the upper-leg light is $\sigma^{-}$ polarized. (c)  Measured lower-leg transmission $T_\mathrm{l}$ with EIT signal. The lower-leg detuning is scanned across the $^{87}$Rb $\ket{5S_{1/2},\text{ }F=2} \rightarrow \ket{6P_{3/2}, \text{ }F'=3}$ hyperfine transition while the upper-leg beam is resonant with the $^{87}$Rb $\ket{6P_{3/2}, \text{ }F'=3} \rightarrow \ket{30S_{1/2}}$ transition. Inset denotes the EIT peak. The vapor cell is held at $89^{\circ}$C. TPAT beam parameters: upper-leg beam power $P_{\mathrm{u}} = 0.3$~mW, upper-leg waist $w_{0\mathrm{u}} = 200$~$\mu$m, lower-leg beam power $P_{\mathrm{l}} = 15$~mW, lower-leg beam waist at cell input $w_{0\mathrm{ l}} = 300$~$\mu$m, lower-leg beam waist at cell output $w_{0\mathrm{l}} = 220$~$\mu$m. EIT beam parameters: $P_{\mathrm{u}} = 19$~mW, $w_{0\mathrm{u}} = 200$~$\mu$m, $P_{\mathrm{l}} = 0.7$~mW, cell input $w_{0\mathrm{l}} = 300$~$\mu$m, cell output $w_{0\mathrm{l}} = 220$~$\mu$m. For EIT, the lower-leg intensity is $I_{l} = 9.9I_{\mathrm{sat}}$, where $I_{\mathrm{sat}}$ is the saturation intensity for the $\ket{5S_{1/2},\text{ } F=2}\rightarrow \ket{6P_{3/2}, \text{ }F'=3} $ transition assuming an atomic population equally distributed in the ground state $m_F$ sublevels and a $\sigma^+$ polarized light field.}\label{fig:demospec}
\end{figure}

We experimentally examine both the EIT and TPAT feature in a Rydberg ladder scheme via the $\ket{5S_{1/2}} \rightarrow \ket{6P_{3/2}} \rightarrow \ket{nS_{1/2}}$ transition with the set-up shown in Fig.\ \ref{fig:demospec}(a). A laser with an emission 420~nm in wavelength (lower-leg) and a laser with an emission 1012~nm to 1026~nm (coupling to Rydberg levels between 30 and 80) in wavelength (upper-leg) generate beams that counter-propagate through a $^{87}$Rb  vapor cell that is 7.5 cm in length and heated to a temperature as high as 96$^\circ$C via a copper-nichrome heating element. The lower-leg light is right-hand circularly polarized and the upper-leg light is left-hand circularly polarized. The transmission of each beam is monitored on a separate photodiode. We focus the lower-leg beam down in the cell so as to homogenize the beam intensity as it is absorbed by the optically dense atomic gas within the cell. The beams are shaped so that the upper-leg beam is enveloped by the lower-leg beam, maximizing the efficiency of two-photon excitation. Transmission of the upper-leg and lower-leg light is probed to obtain the TPAT and EIT features, respectively, revealing the spectroscopic signals shown in Fig.\ \ref{fig:demospec}(b) and (c) for a Rydberg level of $n = 30$.  We additionally plot simulated velocity-averaged TPAT transmission spectra using our experimental parameters for both an unpolarized atomic ensemble (dashed green) and a fully spin-polarized ensemble prepared in the $\ket{5S_{1/2}, \text{ }F=2, \text{ }m_F=2}$ state (dashed purple). The experimentally measured TPAT spectrum (solid red) lies between these two simulated limits. We attribute this behavior to imperfect optical pumping by the circularly polarized lower-leg light field, which preferentially transfers population into higher $m_F$ sublevels but does not produce a fully polarized ensemble. As a result, the experimental population distribution interpolates between the unpolarized and fully polarized cases assumed in the simulations.

We compare the transmissive contrast of the two signals. For TPAT, we define this contrast to be the dip in the upper-leg transmission induced by the two-photon process normalized to the transmission when the upper-leg light is off two-photon resonance. For EIT, we define this contrast to be the increase in the lower-leg transmission induced by the two-photon process normalized to the reduction in transmission induced solely by single-photon absorption of the lower-leg light. We obtain a TPAT contrast of approximately $2 \times 10^{-3}$ and an EIT contrast of approximately $6 \times 10^{-3}$. The comparatively lower value of the TPAT contrast is due to the weak oscillator strength of the upper-leg transition, yielding a relatively small amount of absorption of the incident upper-leg light.

\begin{figure}
	\centering
	\includegraphics[width=\linewidth]{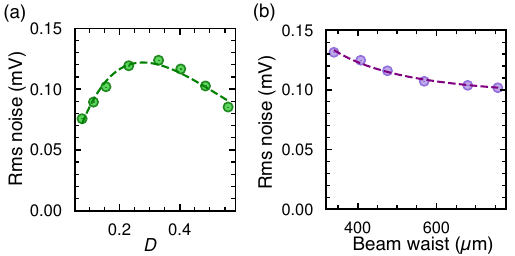}
	\caption{ (a) Rms noise of the lower-leg single-photon absorption over various optical densities. The rms noise is obtained via taking the square root of the integrated noise power spectral density of the photodiode output over a bandwidth of 400~kHz, corresponding to the average transit time of the atoms through the beam. Dashed line corresponds to fit (described in the text). (b) rms noise as a function of beam waist at fixed optical density. Dashed line corresponds to fit (described in the text).  }
	\label{fig:noise_fig}
\end{figure}

We observe substantially larger root-mean-square (rms) voltage fluctuations on the EIT signal than on the TPAT signal. In our setup, the noise power spectral density of the TPAT signal is dominated by residual technical noise, whereas the EIT signal exhibits an approximately threefold increase above the technical noise floor. We find that the primary source of rms noise on the EIT signal is atom number fluctuations on the lower-leg beam transmission. We determine this via first measuring the rms noise of the single-photon absorption background -- which exhibits a noise profile similar to that of the EIT signal -- over a range of optical densities D, observing results that are well-fitted by $\Delta T = \sqrt{(a \sqrt{D} e^{-b D})^2 + c^2}$ where $T$ is the transmission, $a$ and $b$ are additional fitting parameters, and the technical noise floor $c$ is added in quadrature (see Fig.\ \ref{fig:noise_fig}(a)). This functional form follows from differentiating the Beer–Lambert law, $\Delta T = \Delta D\, \frac{dT}{dD}$, with optical-density fluctuations $\Delta D \propto \sqrt{D}$, as expected for Poisson-distributed atom number fluctuations.

We then confirm that these optical density fluctuations arise from atom number fluctuations by expanding the waist of the lower-leg beam to variable sizes (by the $1/e^2$ intensity beam-waist radius $w$) and examining the subsequent change in rms noise. Poisson-distributed atom number fluctuations yield an rms noise that scales as $1/\sqrt{N}$, where $N$ is the number of atoms being probed and is proportional to $w^2$. Adding the technical noise in quadrature, we obtain the fitting function $V_{\mathrm{rms}} = \sqrt{a^2/w^2 + b^2}$. We fit this function to our measurements, obtaining good agreement (see Fig.\ \ref{fig:noise_fig}(b)), and thus conclude that the rms noise on the EIT signal primarily arises from atom number fluctuations. It should be noted that though the atom number fluctuations can be reduced by expanding the waist of the lower-leg beam, this modification greatly increases the required upper-leg optical power to attain a comparable EIT signal amplitude. 

We believe the reason for the relatively lower rms noise on the upper-leg spectroscopic signal is two-fold. First, for single-photon absorption of the lower-leg beam, only a narrow range of atomic velocity classes within the natural linewidth of the intermediate state contributes to the signal, reducing the number of atoms participating in the transition and increasing the sensitivity of the feature to atom number fluctuations. In contrast, the TPAT feature arises from the symmetric Autler–Townes splitting induced by the strong lower-leg field, which produces a turning point in the velocity–detuning spectrum (see dashed pink line in Fig.\ref{fig:sys_theory}(c)). Near this turning point, many atomic velocity classes map onto nearly the same upper-leg light detuning, yielding a stronger two-photon signal and reduced noise from population fluctuations. Second, the amplitude of the EIT window is roughly $2 \times 10^{-2}$ the magnitude of the single-photon absorption dip. Consequently, noise associated with the large single-photon absorption feature is transferred onto the comparatively small EIT signal, disproportionately reducing its resolution. There is no similar competing mechanism for the TPAT resonance.

\section{Rydberg Resonance Location}

We benchmark the signal quality at increasing Rydberg levels by examining the raw signal-to-noise ratio, the signal amplitude divided by the off-resonant background rms noise, and an idealized signal-to-noise ratio, the signal amplitude divided by the rms noise with the detector noise subtracted.

\begin{figure}
	\centering
	\includegraphics{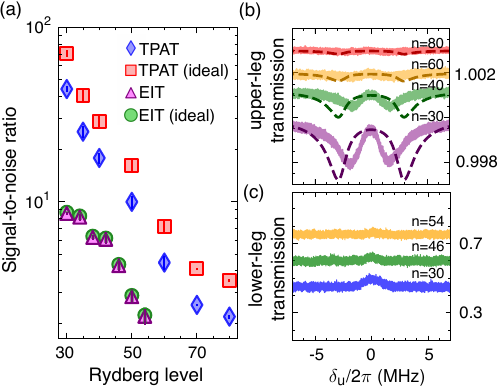}
	\caption{(a) Measured and ideal signal-to-noise ratio as a function of Rydberg level for TPAT and EIT features. The measurement bandwidth is 1 MHz and the vapor cell is held at 95$^\circ$. (b) Measured TPAT transmission for various $n$ levels. Transmission values for higher $n$ are offset for visual clarity. Dashed lines correspond to simulation results with the experimental parameters when all atoms are initialized in the $\ket{5S_{1/2},\text{ } F=2,\text{ }m_F=2}$, the lower-leg light is $\sigma^{+}$ polarized, and the upper-leg light is $\sigma^{-}$ polarized. (c) Measured EIT feature traces for various $n$ levels. Transmission values for higher $n$ levels are offset for visual clarity. TPAT Beam parameters: $P_{\mathrm{u}} = 0.4$~mW, $w_{0\mathrm{u}} = 200$~$\mu$m, $P_{\mathrm{l}} = 19.5$~mW, cell input $w_{0\mathrm{l}} = 300$~$\mu$m, cell output $w_{0\mathrm{l}} = 220$~$\mu$m. EIT Beam Parameters: $P_{\mathrm{u}} = 21$~mW, $w_{0 \mathrm{u}} = 200$~$\mu$m, $P_{\mathrm{l}} = 0.8$~mW, input $w_{0\mathrm{l}} = 300$~$\mu$m, cell output $w_{0\mathrm{l}} = 220$~$\mu$m.}

	\label{fig:scan_n}
\end{figure}

We tune the upper-leg laser to scan across Rydberg resonances for increasing principal quantum numbers $n$, observing the results shown in Fig.\ \ref{fig:scan_n}(a). Raw plots of of the TPAT and EIT feature for increasing $n$ level are shown in Fig.\ \ref{fig:scan_n}(b) and (c), respectively. 

We are able to resolve the TPAT signal up to a Rydberg level of $n=80$, whereas the EIT signal is no longer visible after $n=54$. We attribute the comparatively poor raw signal-to-noise ratio of the EIT signal to the rms noise induced by atom number fluctuations previously discussed in Section \ref{sec:setup}.

Moreover, the EIT rms noise is predominantly limited by these atom number fluctuations, and thus subtracting technical noise results in negligible improvement in the signal-to-noise radio. Contrary to this, a significant fraction of the TPAT rms noise in our setup is technical, and the gap between the raw and ideal noise levels suggests that the signal amplitude can be substantially improved with lower-noise detectors and electronics.

The signal-to-noise ratio for each spectroscopic signal decreases with increasing $n$ for similar reasons. For the EIT feature, higher $n$ level reduces the oscillator strength coupling the intermediate state to the Rydberg state, resulting in reduced excitations to the Rydberg state. This, along with longer Rydberg lifetimes at higher $n$, lowers the population transferred to the dark state and lowers the magnitude of the EIT feature. For TPAT, the decrease in the oscillator strength with increasing $n$ results in fewer intermediate state atoms being transferred to the upper Rydberg state and thus a weakened absorption of the upper-leg beam.

Thus, while the EIT signal can be enhanced by increasing the upper-leg Rabi frequency to drive a larger population into the Rydberg state, the TPAT signal is primarily improved by increasing the optical density on the lower-leg transition, thereby increasing the amount of atoms in the intermediate state. This can be achieved by raising the vapor-cell temperature, employing a multi-pass geometry, or increasing the lower-leg Rabi frequency. Additionally, the use of a re-pumping beam that returns atoms from the dark $\ket{5S_{1/2}, \text{ }F=1}$ manifold to the bright  $\ket{5S_{1/2}, \text{ }F=2}$ manifold, as well as state preparation in the $\ket{5S_{1/2}, \text{ }F=2, m_F=2}$ state can further boost transfer to the intermediate state. However, there is a limit on the signal-to-noise ratio of the TPAT resonance, as the absorption of the lower-leg beam eventually saturates and no further atoms can be transferred to the intermediate state. 

\section{Generating an Error Signal}

\begin{figure}
	\centering
	\includegraphics{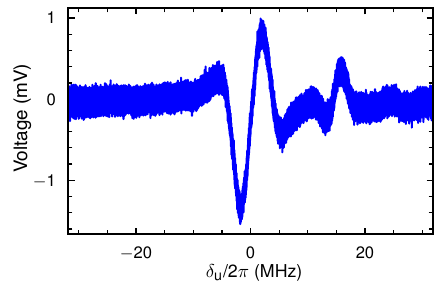}
	\caption{Measured error signal from modulation transfer spectroscopy with a bandwidth of 1~MHz for the Rydberg state $n=30$ at vapor cell temperature $88^\circ$C. Beam parameters: $P_{\mathrm{u}} = 0.5$~mW, $w_{0\mathrm{u}} = 200$~$\mu$m, $P_{\mathrm{l}} = 18$~mW, at the cell input $w_{0\mathrm{l}} = 300$~$\mu$m, at the cell output $w_{0\mathrm{l}} = 220$~$\mu$m. }
	\label{fig:error_sig}
\end{figure}

The comparatively lower noise of the TPAT signal makes it suitable for generating an error signal that can be used to lock the laser driving the upper-leg transition. We demonstrate this via modulation transfer spectroscopy, applying a 300~kHz modulation on the frequency of the laser driving the lower-leg transition and demodulating the transmission of the upper-leg light at the same frequency. Doing so yields the error signal shown in Fig.\ \ref{fig:error_sig}. From the slope and rms noise of the error signal near $\delta_{\mathrm{u}} = 0$, measured to be 0.66~mV/MHz and 0.04~mV, respectively, we infer a frequency noise sensitivity of $61\text{~} \mathrm{Hz}/\sqrt{\mathrm{Hz}}$ for a measurement bandwidth of 1~MHz.

The central transmission maximum of the TPAT feature marks the location of the two-photon resonance, understood here as a peak in transmission arising from the symmetric overlap of Autler–Townes–split dressed states rather than a true eigenstate resonance. When only the upper-leg laser is locked to this signal, its frequency will vary with changes in the frequency of the lower-leg laser.  Upon additionally locking the lower-leg laser to a fixed reference -- such as the one-photon lower-leg resonance, e.g., using saturated-absorption spectroscopy \cite{glaser_sas} -- the upper-leg laser will now also be locked to the corresponding single-photon atomic resonance. 

We note that the capture range of the lock can be dynamically tuned via adjustment of the lower-leg laser intensity without altering the lock frequency set point, as the Rabi frequency of the lower-leg light tunes the width of the avoided crossing in the TPAT spectroscopy signal. We find that we can increase the lock capture range over 8~MHz without any significant change in the error signal slope.

We note that the error signal is broad compared to error signals that can be obtained from commonly used optical frequency references such as ultra-low-expansion cavities. Although the natural linewidth of the intermediate state and the splitting of the Autler-Townes features limits the linewidth of the locked laser, the TPAT atomic signal remains effective for correcting slow drifts. Additionally it may be a lower cost and lower maintenance alternative to ultra-low-expansion cavity-based locking methods, which require periodic recalibration due to cavity mirror drift \cite{lahaye_2024,Ito_17}. Locking to the TPAT atomic signal is particularly suitable for classes of lasers, such as whispering gallery mode resonator-based devices \cite{Toropov2021}, which are intrinsically narrow linewidth but still require a locking mechanism to correct for slow drift.

\section{Conclusion}
We have experimentally investigated the TPAT spectroscopic feature obtained via probing the transmission of the upper-leg laser in a Rydberg ladder system with inverted wavelengths.  We find that the TPAT feature is advantageous over the EIT feature for locating Rydberg states due to lower noise from atom number fluctuations. We believe that the TPAT feature is thus useful in inverted two-step Rydberg excitations for calibrating laser systems when the resonance frequencies are well known or for experimental location of Rydberg resonances in new atomic species.

We have demonstrated that the TPAT feature can also be used to generate an error signal to lock the frequency of the laser driving the upper excitation. Such a locking scheme, when paired with lasers with intrinsically low linewidths that are susceptible to slow drift, represents an additional option for locking the laser that drives the upper transition of the two-step excitation in inverted three-level schemes.

\section{Acknowledgements}

We thank A. Urvoy and M. B. J. Leibovitch for helpful discussions and N. Vilas for providing review and insightful comments on this manuscript. We acknowledge support from the AFOSR (Grant No.\ FA9550-1910328), from ARO through the MURI program (Grant No.\ W911NF-20-1-0136) and the DURIP program (Grant No.\ W911NF2310244), from DARPA (Grant No.\ W911NF2010090), from the NSF through the QLCI program (Grant No.\ OMA-2016245) and the MRI program (Grant No.\ OMA-2216201), and from the U.S. Department of Energy, Office of Science, National Quantum Information Science Research Centers, Quantum Systems Accelerator.
J.H. acknowledges support from the Department of Defense through the National Defense Science and Engineering Graduate (NDSEG) Fellowship Program.

\bibliography{Main.bib}
\end{document}